\documentclass[a4paper, 11pt]{article}
\usepackage{graphicx}
\usepackage{amsmath}
\usepackage{amsfonts}
\usepackage{natbib}
\usepackage{amssymb}
\usepackage{rotating}

\usepackage{setspace}

\newcommand{\var}[1]{\ensuremath{\mathrm{Var}[#1]}}

\newcommand{\bd}[1]{\ensuremath{\mbox{\boldmath $#1$}}}

\begin{document}

\doublespacing

\title{Cluster detection and risk estimation for spatio-temporal health data}

\author{Duncan Lee$^1$ and Andrew Lawson$^2$\\
$^1$ School of Mathematics and Statistics, University of Glasgow\\
$^2$ Division of Biostatistics and Bioinformatics, Department of Public Health Sciences, \\Medical University of South Carolina}

\maketitle

\begin{abstract}
{In epidemiological disease mapping one aims to estimate the spatio-temporal pattern in disease risk and identify high-risk clusters, allowing health interventions to be appropriately targeted. Bayesian spatio-temporal models are used to estimate smoothed risk surfaces, but this is contrary to the aim of identifying  groups of areal  units that exhibit elevated risks compared with their neighbours. Therefore, in this paper we  propose a new Bayesian hierarchical modelling approach for simultaneously estimating disease risk and identifying high-risk clusters in space and time. Inference for this model is based on Markov chain Monte Carlo simulation, using the freely available R package \emph{CARBayesST} that has been developed in conjunction with this paper. Our methodology is motivated by two case studies, the first of which assesses if there is a relationship between Public health Districts and colon cancer clusters in Georgia, while the second looks at the impact of the smoking ban in public places in England on cardiovascular disease clusters.}

\textbf{Bayesian modelling, Cluster detection, Software, Spatio-temporal health mapping}
\end{abstract}

\section{Introduction}
The risk of disease varies in both space and time, as a result of variation in environment exposures, risk inducing behaviors and public health policy. Modelling and mapping  this spatio-temporal variation in disease risk has substantial health, financial and societal benefits. The public health benefits include the ability to estimate temporal disease trends and detect spatial clusters of areas exhibiting high risks, enabling timely and targeted interventions for disease outbreaks.  The financial benefit is the ability to forecast future disease burdens, yielding predictions of the required size and geographical spread of future health resources.  Disease maps also enable policymakers to understand the nature and scale of health inequalities between different social groups, which lead to social injustice and may undermine social cohesion. The scale of such health inequalities can be substantial, with a 2014 Office for National Statistics report finding a differential of nineteen years in average healthy life expectancy  between the richest and poorest neighbourhoods in England (\citealp{ons2014}). \\

Maps of raw disease rates are routinely produced by health agencies such as Public Health England (PHE) and the Centers for Disease Control and Prevention (CDC), but they are contaminated by random noise and do not provide a natural framework for inference. Therefore, Bayesian spatio-temporal models for risk estimation have been proposed by \cite{bernardinelli1995}, \cite{knorrheld2000} and \cite{lawsonBDM13}, while scan statistics have been proposed for cluster detection by \cite{kulldorff2005}. However, these modelling approaches have fundamentally different goals, as the former aims to use the spatio-temporal autocorrelation  present in the data to estimate smoothed maps of disease risk, while the latter aims to identify clusters of areal units that exhibit elevated risks compared with neighbouring areas. Risk estimation and cluster identification are both epidemiologically important goals, and it would be preferable if both could be addressed in a unified modelling framework, rather than doing separate analyses, possibly using different software packages, which make different assumptions about the data.\\

A two-stage approach for addressing both aims was proposed by \cite{charras2013} in a purely spatial context, which applies a clustering algorithm to the risk surface estimated from a spatial smoothing model. Attempting to identify clusters (i.e. discontinuities between neighbouring areas) from a spatially smoothed risk surface is inherently problematic, and \cite{anderson2014} show that such an approach does not lead to good cluster recovery. Alternatively, clusters could be identified based on exceedence probabilities (\cite{richardson2004}), but as tail area probabilities these are sensitive to the model chosen and require the user to specify a threshold level for exceedences. Alternatively,  \cite{gangnon2000}, \cite{knorrheld2000b}, \cite{green2002}, \cite{forbes2013}, \cite{wakefield2013} and \cite{anderson2014} have proposed  an integrated modelling approach to risk estimation and cluster identification in a purely spatial context, while only \cite{choiENV11}, \cite{lawsonJABES12} and  \cite{li2012} have extended these models to the spatio-temporal domain. However, the latter have focused on detecting shared latent structures and unusual temporal trends, and an integrated modelling framework for spatio-temporal risk estimation and cluster detection is yet to be proposed.\\

The contribution of this paper is to fill this gap, by proposing a novel Bayesian spatio-temporal model for cluster detection and risk estimation.  The model is able to detect clusters dynamically, so that both cluster membership  and their average risk levels can evolve over time. Inference is based on Markov chain Monte Carlo (McMC) simulation, and unlike most existing models for spatio-temporal disease mapping, our methodology can be implemented in freely available software, making this research reproducible. This software is the R package \emph{CARBayesST}, which was  developed in conjunction with this paper. The methodology developed here is motivated by two case studies, and the motivating questions for both analyses focus on the relationship between public health policy and disease clustering. The background and motivation for these case studies are outlined in Section 2, together with a brief review of spatio-temporal disease mapping. Section 3 presents our methodological contribution, and details the software that has been developed. Section 4 quantifies the performance of our methodology by simulation, while the results from the two case studies are presented in Section 5. Finally, Section 6 concludes the paper.

\section{Background and motivation}

\subsection{Spatio-temporal disease mapping}
Disease data ($y_{it},e_{it}$) denote the observed and expected numbers of disease cases across $i=1,\dots,N$ contiguous areal units during  $t=1,\ldots,T$ time periods. The latter, $e_{it}$, is computed using internal or external standardisation, and accounts for the heterogeneous population sizes and demographic structures across the $N$ areal units. The same set of age and sex specific disease rates are used in the standardisation for all time periods, because any estimated temporal trend in disease risk is then relative to a common baseline. The Poisson model $y_{it}|e_{it},\theta_{it}~\sim~\mbox{Poisson}(e_{it}\theta_{it})$ is commonly assumed for the disease counts $y_{it}$ in the spatio-temporal disease mapping literature, where $\theta_{it}$ is the risk of disease in areal unit $i$ during time period $t$ relative to $e_{it}$. Log-linear models are typically specified for $\theta_{it}$, and the first model to be proposed in the literature was \cite{bernardinelli1995}, who represented $\{\theta_{it}\}$ by a  set of spatially varying linear time trends. However, linear trends are not always appropriate, and the most popular model in the literature is the main effects and interaction decomposition  proposed by \cite{knorrheld2000}:

\begin{eqnarray}
\ln(\theta_{it})&=&\beta + \phi_i + \theta_i+ \alpha_t + \delta_t  + \gamma_{it},\label{knorrheld}\\
\phi_i|\bd{\phi}_{-i},W,\tau^2_{\phi}&\sim&\mbox{N}\left(\frac{\sum_{j=1}^Nw_{ij}\phi_j}{\sum_{j=1}^Nw_{ij}}, \frac{\tau^2_{\phi}}{\sum_{j=1}^Nw_{ij}}\right),\nonumber\\
\theta_i|\tau^2_{\theta} &\sim&\mbox{N}(0, \tau^2_{\theta}),\nonumber\\
\alpha_t|\alpha_{t-1},\tau^2_{\alpha}&\sim&\mbox{N}(\alpha_{t-1}, \tau^2_{\alpha}),\nonumber\\
\delta_t |\tau^2_{\delta}&\sim&\mbox{N}(0, \tau^2_{\delta}),\nonumber\\
\gamma_{it}|\bd{\gamma}_{-it}, \tau^2_{\gamma}&\sim&\mbox{N}(m_{it}, \tau^2_{\gamma} v_{it}).\nonumber\end{eqnarray}

The spatial ($\phi_i+\theta_i$) and temporal ($\alpha_t+\delta_t$) main effects are a convolution of independent ($\theta_i, \delta_t$) and autocorrelated ($\phi_i, \alpha_t$) processes. Spatial autocorrelation is modelled by an intrinsic  conditional autoregressive (ICAR, \citealp{besag1991}) model with a binary $N\times N$ neighbourhood matrix $W$, where $w_{ij}=1$ if areal units $(i,j)$ share a common border and is zero otherwise. Thus the conditional expectation is the mean of the random effects in neighbouring areal units, inducing spatial smoothness into $\phi_{i}$. Temporal autocorrelation is modelled by a first order random walk, where the conditional expectation of $\alpha_t$ is $\alpha_{t-1}$. Four different models were proposed by \cite{knorrheld2000} for the interaction term $\gamma_{it}$, which respectively assume independence, purely spatial autocorrelation, purely temporal autocorrelation and spatio-temporal autocorrelation. The first of these has $(m_{it}=0, v_{it}=1)$, and further details are given by \cite{knorrheld2000}.  Weakly informative Inverse-gamma and Gaussian priors can be specified for the variance parameters  ($\tau^2_{\phi}, \tau^2_{\theta}, \tau^2_{\alpha}, \tau^2_{\delta}, \tau^2_{\gamma}$) and intercept term ($\beta$) respectively.\\

Another popular approach is to allow the spatially smoothed log-risk surface to evolve over time via an autoregressive process, and a number of models of this type have been proposed (see for example \cite{mugglin2002}, \cite{ugarte2012} and  \cite{rushworth2014}). Here we outline the model of \cite{rushworth2014}, which is given by:

\begin{eqnarray}
 \ln (\theta_{it}) & = & \beta + \phi_{it},\label{sste}\\
 \bd{\phi}_1=(\phi_{11},\ldots,\phi_{N1}) &  \sim & \mbox{N}\left(\bd{0}, \hspace{0.1cm}\tau^2 Q(W, \rho)^{-1} \right),\nonumber\\
\bd{\phi}_t|\bd{\phi}_{t-1} &  \sim &\mbox{N}\left(\gamma \bd{\phi}_{t-1}, \hspace{0.1cm}\tau^2 Q(W, \rho)^{-1} \right)  \text{   for   } t=2,\dots,T,\nonumber \\
 \gamma,\rho & \sim & \mbox{Uniform}(0,1).\nonumber
 \end{eqnarray}

Here $\beta$ is again an intercept term and $\bd{\phi}_t=(\phi_{1t},\ldots,\phi_{Nt})$ represents the spatial pattern in the log-risk surface at time period $t$. Temporal autocorrelation between the log-risk surfaces is controlled by the autoregressive parameter $\gamma$, while spatial autocorrelation is induced via the precision matrix $Q(W, \rho)= \rho [\textrm{diag}(W\textbf{1}) - W ] + (1 - \rho) \textrm{I}$, where $\textbf{1}$ is an $N\times 1$ vector of ones and $\textrm{I}$ is the $N \times N$ identity matrix. This precision matrix corresponds to the conditional autoregressive model proposed by \cite{leroux1999}, which for time period 1 has the full conditional decomposition:

\begin{equation}
\phi_{i1}|\bd{\phi}_{-i1}, \rho, \tau^{2}, W \sim  \mbox{N}\left(\frac{\rho \sum_{j=1}^{N} w_{ij}\phi_{j1}}{\rho\sum_{j=1}^{N}w_{ij} + 1 - \rho}, \frac{\tau^2}{\rho\sum_{j=1}^{N}w_{ij} + 1 - \rho}\right).\label{leroux}
\end{equation}

The level of spatial smoothing is controlled by $\rho$, and if $\rho=1$ equation (\ref{leroux}) reduces to the ICAR model while if $\rho=0$ the random effects have identical and independent normal prior distributions. Although (\ref{knorrheld}) and (\ref{leroux}) are appropriate models for risk estimation, neither have an inbuilt mechanism for identifying clusters of areas exhibiting elevated disease risks. Thus in Section 3 we extend (\ref{sste}) to allow for such cluster detection, but before that we present our two motivating case studies.

\subsection{Case study 1 - The relationship between Public Health Districts (PHD) in Georgia and disease clustering}
\begin{figure}
\centering\caption{Maps showing the SIR for colon cancer incidence in Georgia in 2001 and 2010 (top panel) and circulatory disease in England in 2003 quarter 1 (Q1) and 2011 quarter 4 (Q4) (bottom panel). For the latter the Easting and Northing coordinates are in kilometres.}\label{figure SIRplot}
\begin{picture}(10,21)
\put(-9,0){\scalebox{0.8}{\includegraphics{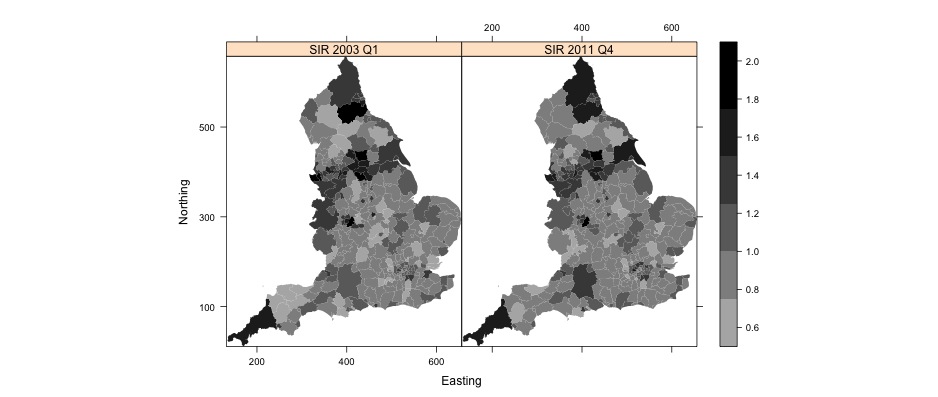}}}
\put(-8,11){\scalebox{0.95}{\includegraphics{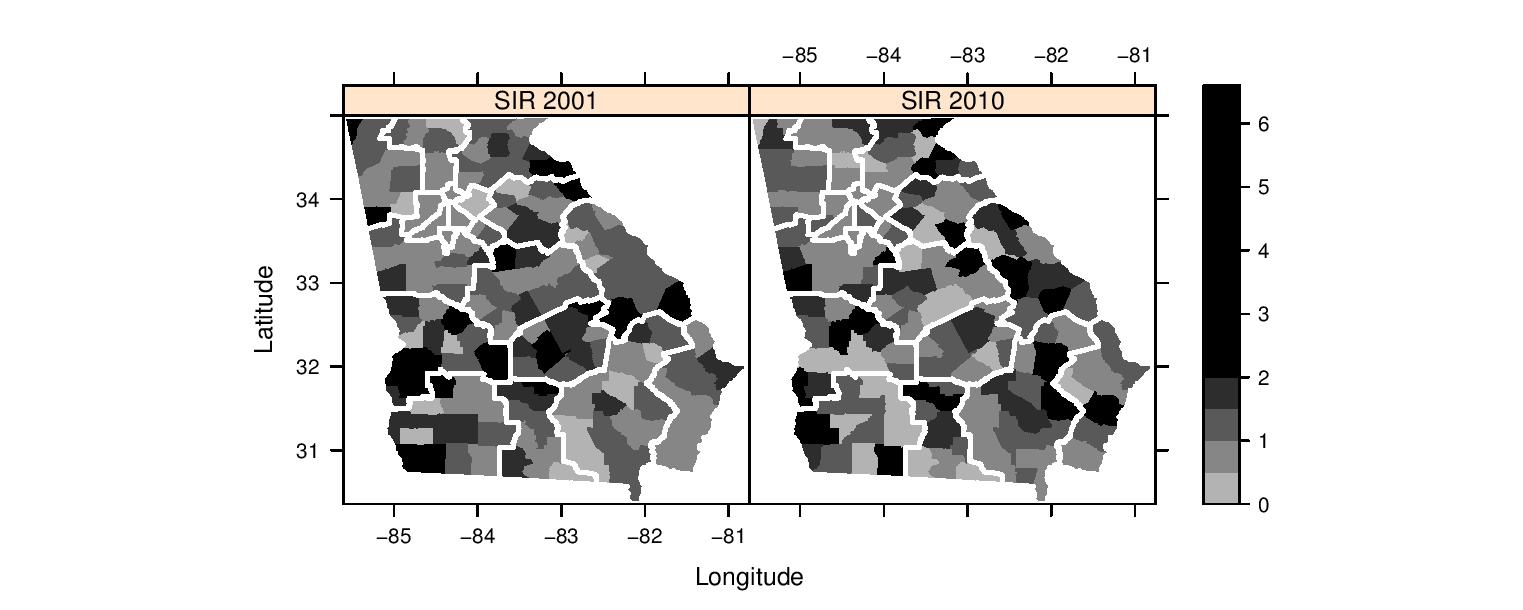}}}
\end{picture}
\end{figure}

The state of Georgia, USA contains $N=159$ counties, which for the provision of healthcare are grouped into 18 Public Health Districts containing between 1 and 16 counties  (for details see \emph{http://dph.georgia.gov/public-health-districts}). The motivating question for this analysis is whether disease risk clusters by PHD, which may be indicative of differential levels of health care across the 18 PHD. We test this hypothesis using yearly counts of colon cancer incidence  (International Classification of Disease tenth revision (ICD-10) code C18) between 2001 and 2010, and maps of the standardised incidence ratio (SIR$_{it}=y_{it}/e_{it}$) for the first (2001) and last (2010) years of the study period are displayed in the top row of Figure \ref{figure SIRplot}. The white lines on the maps depict the 18 Public Health Districts in Georgia, and no clear relationship between SIR and PHD exists for either of the two years. The SIRs also exhibit substantial skewness, with little in the way of a spatially consistent pattern between the two years. This lack of spatial consistency occurs in all years, as the maximum pairwise correlations in SIR in different years is 0.47.  An exploratory analysis of the relationship between SIR and PHD  shows weak evidence for a relationship. A two-way analysis of variance incorporating the effects of year and PHD shows a statistically significant effect of PHD on SIR (p-value $1.098\times 10^{-12}$), but this only explains 5.8$\%$ of the variation in the SIR data. Therefore, the spatio-temporal clustering model proposed in this paper will be applied to these data in Section 5 to provide a more conclusive answer to this question.

\subsection{Case study 2 - The impact of the smoking ban in England on circulatory disease clustering}
A ban on smoking in indoor public places such as restaurants, bars, etc was introduced in England on 1st July 2007,  following similar bans in Ireland (2004) and Scotland (2005). The goal of this case study is to assess whether the English ban has had any effect in reducing circulatory disease risk, and whether any such reduction applies equally across communities with different baseline risk levels. Thus the clustering question here is the extent to which the clusters identified are consistent over time, and whether the cluster specific mean risk levels have reduced since July 2007. To answer these questions we use quarterly count data quantifying the numbers of admissions to hospital  due to circulatory disease (ICD-10 codes I00-I99) for the set of $N=323$ local authorities in mainland England between 2003 to 2011. The expected counts for all quarters were computed using age and sex specific disease rates for England in 2003, the first year of the study. The SIRs for the first (2003 quarter 1) and last (2011 quarter 4) time periods are displayed in the bottom panel of Figure \ref{figure SIRplot}, and show definite clusters of high-risk areas in Greater Manchester / West Yorkshire (mid to north of the study region) and Tyneside (far north east of the study region). These clusters appear to be temporally stable, as the exploratory analysis of these data shows a mean pairwise correlation of 0.95 between the spatial SIR surfaces across different time periods. The exploratory analysis also suggests no clear decline in risk on average since 2007, although this global pattern may hide cluster specific patterns and the analysis in Section 5 of the main paper will address this issue.

\section{Methodology}
This section describes the general Bayesian spatio-temporal cluster detection and risk estimation model developed in this paper, as well as outlining the software that has been developed to allow others to apply it to their own data.

\subsection{Likelihood model}
The disease data $(y_{it},e_{it})$ are modelled with the following Poisson log-linear model:

\begin{eqnarray}
y_{it}|e_{it},\theta_{it}&\sim&\mbox{Poisson}(e_{it}\theta_{it})~~~~\mbox{for }i=1,\ldots,N \mbox{  and  }t=1,\ldots,T,\label{likemodel}\\
\ln(\theta_{it})&=&\lambda_{t,Z_{it}} + \phi_{it},\nonumber
\end{eqnarray}

which uses the same Poisson log-linear structure as the majority of the existing spatio-temporal disease mapping literature. The spatio-temporal pattern in the log-risk surface is modelled as a linear combination of a smooth space-time component $\phi_{it}$ and a piecewise constant clustering or intercept component $\lambda_{t,Z_{it}}$, which allows disease risk to vary smoothly between neighbouring areas within the same cluster but exhibit a step-change if those areas are in different clusters. Thus if the spatio-temporal risk surface is completely smooth with no clusters then the smoothing component ($\phi_{it}$) will dominate the clustering component ($\lambda_{t,Z_{it}}$), while at the other extreme for a piecewise constant risk surface the clustering component $\lambda_{t,Z_{it}}$ will dominate. A general model for the piecewise constant cluster component $\lambda_{t,Z_{it}}$ is presented in Section 3.2, while potential models for the spatial smoothing component are presented in Section 3.3.

\subsection{Clustering component $\lambda_{t,Z_{it}}$}
The clustering component  for time period $t$ comprises at most $G$ distinct (log) risk levels or classes $(\lambda_{t1},\ldots,\lambda_{tG})$, so that it is essentially a piecewise constant intercept term over the set of $N$ areal units. These risk classes are ordered as $\lambda_{t1}<\lambda_{t2}<\ldots<\lambda_{tG}$, which helps mitigate against the label switching problem common in mixture models. \emph{A-priori}, one would expect the mean risk levels in each class to evolve smoothly over time, which is achieved by putting first order random walk priors on the set $(\lambda_{1j},\ldots,\lambda_{Tj})$ for each risk class $j=1,\ldots,G$. However, the cluster means follow the ordering constraint $\lambda_{t1}<\lambda_{t2}<\ldots<\lambda_{tG}$ across the $G$ groups for each time period, leading to the constrained random walk prior model:

\begin{eqnarray}
\lambda_{tj}|\lambda_{t-1,j}&\sim&\mbox{N}(\lambda_{t-1,j},\sigma^{2})_{I(\lambda_{tj}\in[\lambda_{t,j-1},\lambda_{t,j+1}])}~~~~\mbox{for }j=1,\dots,G, ~~t=2,\ldots,T,\label{clustermodel2}\\
\lambda_{1j}&\sim&\mbox{Uniform}(\lambda_{1,j-1},\lambda_{1,j+1})~~~~\mbox{for }j=1,\dots,G,\nonumber\\
\sigma^{2}&\sim&\mbox{Inverse-Gamma}(a=0.001,b=0.001),\nonumber
\end{eqnarray}

where $I(.)$ denotes an indicator function. At time period one a flat prior is specified on the interval obeying the same ordering constraint, and in the above notation $\lambda_{t0}=-\infty$ and  $\lambda_{t,G+1}=\infty$ for all $t$. \\

 The assignment of areal unit $i$ to  one of the $G$ risk classes during time period $t$ is controlled by the class indicator $Z_{it}\in\{1,\ldots,G\}$, and our choice of prior $f(Z_{it})$ is guided by two considerations. First, one may expect the risk in an area to evolve smoothly in time, which would favour a temporally autocorrelated prior for $Z_{it}$ such as a Markov model. The second consideration relates to how the number of risk classes $G$ is chosen, and here we do not estimate it in the model using a reversible jump McMC algorithm (\citealp{knorrheld2000b}). Instead we adopt a penalty based approach, which fixes $G$ to be overly large and then uses the prior for $Z_{it}$ to penalise values in the extreme risk classes. Thus $\{Z_{it}\}$ are allowed to take values in the set $\{1,\ldots,G\}$ but are not forced to completely cover the set, meaning that some of these classes may be empty. These two considerations suggest a Markov decomposition for the prior of $\mathbf{Z}=\{Z_{it}|i=1,\ldots,N, t=1,\ldots,T\}$ of the form:
  
\begin{equation}
f(\mathbf{Z})~=~\prod_{i=1}^{N}\left[f(Z_{i1})\prod_{t=2}^{T}f(Z_{it}|Z_{i,t-1})\right].\label{markovmodel1}
\end{equation}

No spatial autocorrelation is enforced on $\mathbf{Z}$, because two different high-risk clusters could be present on opposite sides of the study region. As each $Z_{it}\in\{1,\ldots,G\}$ we propose a discrete prior distribution that combines a Markov component with a penalty component and is given by:

\begin{eqnarray}
f(Z_{it}|Z_{i,t-1})&=&\frac{\exp(-\alpha(Z_{it}-Z_{i,t-1})^{2} -\delta(Z_{it}-G^{*})^{2})}{\sum_{r=1}^{G}\exp(-\alpha(r-Z_{i,t-1})^{2}-\delta(r-G^{*})^{2})}~~~~\mbox{for }i=1,\dots,N, ~~t=2,\ldots,T,\label{clustermodel3}\\
f(Z_{i1})&=&\frac{\exp(-\delta(Z_{it}-G^{*})^{2})}{\sum_{r=1}^{G}\exp(-\delta(r-G^{*})^{2})},\nonumber\\
\alpha,\delta&\sim&\mbox{Uniform}(0,M)~~~~\mbox{for }i=1,\dots,N.\nonumber
\end{eqnarray}

Temporal autocorrelation in $\mathbf{Z}$ is controlled by the $-\alpha(Z_{it}-Z_{i,t-1})^{2}$ component, and $\alpha=0$ corresponds to independence of ($Z_{it}, Z_{i,t-1}$). The penalty term $\delta(Z_{it}-G^{*})^{2}$ penalises class indicators $Z_{it}$ towards the middle risk class $G^{*}=(G+1)/2$, so that $\delta=0$ corresponds to having no such smoothing penalty. This penalty term is required because $G$ is fixed to be overly large rather than estimated, as it guards against a potential lack of identifiability in the model. For example, if $G=3$ but the data contain only one risk class (i.e. a background risk level close to one), then the model will fit equally well if all $Z_{it}=1,2,3$ when $\delta=0$. The smoothing parameters $(\alpha, \delta)$ are assigned uniform priors on a large range ($M$ is chosen to be overly large), and are simultaneously estimated in the model.

\subsection{Smoothing component $\phi_{it}$}
The smoothing component $\phi_{it}$ models spatially and temporally autocorrelated variation in the log risk surface, which allows two adjacent areal units and time periods within the same cluster to have similar but not identical disease risks. A number of models are possible to represent this smooth variation, and we consider a range of them here.

\subsubsection{\textbf{Model-1: no smoothing}}
The two components $(\lambda_{t,Z_{it}}, \phi_{it})$ may compete with each other to represent the same variation in the data, potentially leading to disease clusters wrongly being modelled by the smoothing component $\phi_{it}$. Therefore the first alternative we consider is the simplification $\phi_{it}=0$ for all areas and time periods, which assigns all the variation in disease risk to the non-smooth clustering component. 

\subsubsection{\textbf{Model-2: spatio-temporal smoothing}}
The random effects $\{\phi_{it}\}$ can be smoothed in space and time using the model of \cite{rushworth2014} that is described in Section 2, which has the autoregressive decomposition:

\begin{eqnarray}
 \bd{\phi}_1 &  \sim & \mbox{N}\left(\bd{0}, \hspace{0.1cm}\tau^2 Q(W, \rho)^{-1} \right),\label{smoothing model}\\
\bd{\phi}_t|\bd{\phi}_{t-1} &  \sim &\mbox{N}\left(\gamma \bd{\phi}_{t-1}, \hspace{0.1cm}\tau^2 Q(W, \rho)^{-1} \right)  \text{   for   } t=2,\dots,T.\nonumber
 \end{eqnarray}

Here uniform priors on the unit interval are specified for the temporal and spatial autocorrelation parameters ($\gamma, \rho$), and further details are given in Section 2.

\subsubsection{\textbf{Model-3: spatial CAR smoothing}}
Both the mean risk levels $\{\lambda_{tj}\}$ and the class indicators $\{Z_{it}\}$ are temporally autocorrelated, and additional temporal autocorrelation in the smoothing component may be unnecessary. Therefore we consider a simplification of Model 2 where $\gamma=0$, which spatially smoothes the random effects with independent CAR priors for each time period.

\subsubsection{\textbf{Model-4: spatial convolution smoothing}}
Finally, a CAR style structure may not be the best model for the smoothing component, so we also consider a convolution style model similar to \cite{best2005}:

\begin{eqnarray}
\phi_{it}&=&\sum_{j=1}^{N}f(d_{ij}|\rho)X_{jt},\label{convmodel}\\
X_{jt}&\sim&\mbox{N}(0, \tau^{2}),\nonumber\\
\tau^{2}&\sim&\mbox{Inverse-Gamma}(a=0.001,b=0.001),\nonumber\\
\rho&\sim&\mbox{Uniform}(0,P).\nonumber
\end{eqnarray}

Here $f(d_{ij}|\rho)$ is a scaled Gaussian kernel of distance $d_{ij}$ between two area's centroids defined by $f(d_{ij}|\rho)~=~ \exp(-d_{ij}^{2}/(2\rho))/ \sum_{k=1}^{N}\exp(-d_{ik}^{2}/(2\rho))$, except that $f(d_{ii}|\rho)=0$. Here $\rho$ acts as a spatial smoothness parameter, with larger values of $\rho$ leading to spatially smoother surfaces. The upper limit on the prior for $\rho$, $P$, is chosen to be overly large, so that different levels of spatial smoothness can be estimated from the convolution.

\subsection{Model interpretation and software}
An add-on package to the statistical software R (\citealp{R}) called \emph{CARBayesST} has been developed in conjunction with this paper to implement the models discussed here, which is one of the first R packages to implement spatio-temporal models for areal unit data. The package is freely available to download from \emph{http://cran.r-project.org/}, and can fit the model class proposed here as well as the models of \cite{knorrheld2000} (with independent interaction effects) and \cite{rushworth2014}. For the cluster model proposed here the spatial cluster structure in the data is defined by the posterior median of the risk class allocation variable $\{Z_{it}\}$. This is because if two adjacent areas have the same $Z$ value then their risks will either be identical (when $\phi_{it}=0$) or spatially smoothed, meaning they are in the same risk cluster. In contrast, if two adjacent areas $(i,j)$  have different $Z$ values then their risk levels could be substantially different, due to having different intercept parameters $(\lambda_{t,Z_{it}},\lambda_{t,Z_{jt}})$. Thus this corresponds to being in different spatial clusters.

\section{Model assessment via simulation}
This section presents a simulation study, which compares the four different specifications for the  general clustering model proposed in Section 3 with existing disease mapping models. The latter include the main effects and interaction decomposition proposed by \cite{knorrheld2000} with independent interactions, as well as the autoregressive model of \cite{rushworth2014}. Each of the clustering models is fitted with a maximum of $G=5$ risk classes, where as the simulated data exhibit either one or two distinct risk levels.

\subsection{Data generation and study design}
\begin{figure}
\centering\caption{Maps showing example realisations of the SIR surfaces generated in the simulation study. The rows correspond to different values of risk $r$ for the high-risk clusters, while the columns correspond to different disease prevalences $\{e_{it}\}$. High risk clusters are present for $r=2,3$, and their locations are denoted by the black shaded regions in the map on the top right of the figure.}\label{figure simstudy}
\begin{picture}(10,21)
\put(-5,5){\scalebox{0.7}{\includegraphics{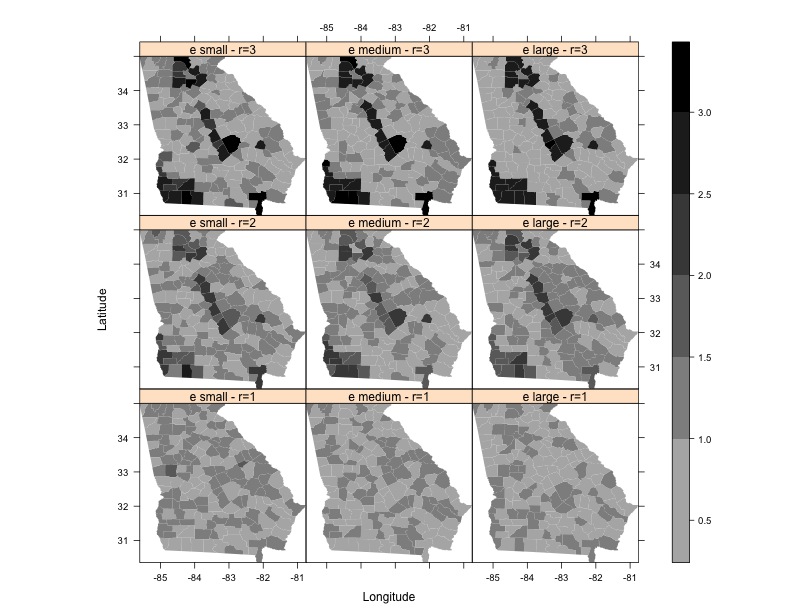}}}
\end{picture}
\end{figure}

Simulated data are generated for the $N=159$ counties comprising the state of Georgia, USA for a period of $T=10$ years, which is the study region for the first case study outlined in Section 2.2. Disease counts are generated from a Poisson log-linear model, where the size of the expected numbers $\{e_{it}\}$ is varied in this study to assess their impact on model performance. The log-risk surface is generated from a multivariate Gaussian distribution, with a piecewise constant mean (for clustering) and a spatially smooth variance matrix. The latter induces smooth spatial variation into the log-risks within a cluster, and is defined by an exponential correlation function based on the Euclidean distance between areal unit centroids. The high-risk clusters present in the simulated data are depicted by the black shaded regions in the top row of Figure \ref{figure simstudy}, which displays sample realisations of the SIR surfaces under the set of scenarios considered here (see below). This cluster structure was chosen to contain clusters of different shapes, including a singleton cluster, a long thin cluster, a block cluster and a cluster with a hole in the middle. We consider five different scenarios in this study, which are outlined below.

\begin{enumerate}
\item \textbf{No clustering} - The null scenario where the risk surface exhibits a smoothly varying baseline risk close to one with no  clustering, as illustrated in the bottom row of Figure \ref{figure simstudy}. This scenario tests whether the methods falsely identify clusters when none are present.

\item \textbf{Medium temporally consistent clustering} - The risk surface contains high-risk clusters with average risks of $r=2$ for all time periods, as illustrated in the middle row of Figure \ref{figure simstudy}. This scenario tests whether the methods can identify clusters whose risk is elevated slightly above the background risk level.

\item \textbf{High temporally consistent clustering} - The risk surface contains high-risk clusters with average risks of $r=3$ for all time periods, as illustrated in the top row of Figure \ref{figure simstudy}. This scenario tests whether the methods can identify clusters whose risk is highly elevated above the background risk level.

\item \textbf{Medium temporally inconsistent clustering} - The risk surface contains no high-risk  clusters during time periods 1 to 3 or periods 8 to 10, and the realisations look like those under scenario 1. However, high-risk clusters with an average risk of $r=2$ appear during time periods 4 to 7, yielding SIR surfaces similar to scenario 2. This scenario tests the models ability to detect slightly elevated clusters that are not temporally consistent.

\item \textbf{High temporally inconsistent clustering} - The risk surface contains no high-risk  clusters during time periods 1 to 3 or periods 8 to 10, and the realisations look like those under scenario 1. However, high-risk clusters with an average risk of $r=3$ appear during time periods 4 to 7, yielding SIR surfaces similar to scenario 3. This scenario tests the models ability to detect clusters with highly elevated risks that are not temporally consistent.
\end{enumerate}

The five scenarios above are repeated with expected numbers of disease cases $e_{it}\in[10,30]$, $[90,110]$, $[190,210]$, which allows us to examine model performance for diseases with different underlying prevalences. SIR surfaces generated under these different disease prevalences are displayed in the columns of Figure \ref{figure simstudy}, and show that as $e_{it}$ increases the SIR values are smoother  and exhibit less random noise due to Poisson sampling variation. Inference for each model is based on 10,000 McMC samples, which were generated following a burn-in period of a further 10,000 samples. Convergence was visually assessed to have been reached after 10,000 samples by viewing trace plots of sample parameters for a number of simulated data sets.\\

Two hundred data sets are generated under each scenario, and model performance is summarised using two metrics. The first is the  root mean square error (RMSE) of the estimated risk surface, which summarises a models ability to accurately estimate the spatio-temporal variation in disease risk. RMSE is computed as $\mbox{RMSE}=\sqrt{\frac{1}{NT}\sum_{t=1}^{T}\sum_{i=1}^{N}(\theta_{it} - \hat{\theta}_{it})^{2}}$, where $\hat{\theta}_{it}$ is the posterior median for $\theta_{it}$. The second metric we use is the Rand Index (\cite{rand1971}), which quantifies a models ability to correctly identify the true cluster structure in the data. It is given by $\mbox{Rand}~=~\frac{a+b}{{NT \choose 2}}$, where $a$ is the number of pairs of risks $(\theta_{it}, \theta_{jr})$ that are classified in the same cluster under both  the true and estimated cluster structures, while $b$ is the number of pairs of risks that are classified in different clusters under both cluster structures. The Rand index thus measures the proportion of agreement between the two cluster structures, with a value of one indicating the structures are identical. For the clustering models, the cluster area $i$ is in during time period $t$ is summarised by the posterior median of $Z_{it}$, while the models of \cite{knorrheld2000} and \cite{rushworth2014} have no such inbuilt clustering mechanism. Therefore for these latter models we implement the posterior classification approach described in \cite{charras2012}, which applies a Gaussian mixture model to the posterior median risk surface to obtain the estimated cluster structure.

\subsection{Results}
The results of this study are displayed in Table \ref{table simulation1}, which compares the models proposed by \cite{knorrheld2000} (with independent interactions, denoted \textbf{Model-KH}) and \cite{rushworth2014} (denoted \textbf{Model-RLM}) with the four  different variants of the model proposed in Section 3 (denoted \textbf{Model-1} to \textbf{Model-4}). The top panel of the table displays the RMSE of the estimated risk surface while the bottom panel displays the Rand index quantifying a model's clustering ability, and in both cases the mean values over the 200 simulated data sets are presented. The table shows that in the absence of high-risk clusters (scenario 1) \textbf{Model-RLM} and \textbf{Model-2} perform best in terms of RMSE, as they utilise the same spatio-temporal smoothing component. In contrast, when a risk surface contains high-risk clusters (scenarios 2 to 5) but the disease prevalence is low (small $\{e_{it}\}$) \textbf{Model-KH}, \textbf{Model-RLM} and \textbf{Model-2} perform best, while the other clustering models estimate the risk surface much less accurately.  However, as $\{e_{it}\}$ increases cluster models 2-4 outperform the global smoothing models (\textbf{Model-KH}, \textbf{Model-RLM}), although no one variant from models 2-4 is universally superior. Therefore, \textbf{Model-2} performs best overall in terms of RMSE across the range of scenarios, having the lowest or close to lowest RMSE values in 10 of the 15 scenarios considered. Therefore in general, fitting a model that explicitly accounts for high-risk clusters will lead to improved estimation performance compared with existing models such as \textbf{Model-KH}  and \textbf{Model-RLM} that do not account for such clustering. \\

The Rand index results in the bottom panel of Table \ref{table simulation1} show that applying a clustering algorithm to a smoothed posterior risk surface generally performs poorly, and should not be used for cluster detection in this context. This is particularly evident in scenario 1, where the Rand indices from \textbf{Model-KH} and \textbf{Model-RLM} indicate that clustering the estimated risk surfaces from these  models have resulted in large numbers of false positives being identified.  The results for the clustering models proposed here show that the Rand indices increase (better cluster identification) as $e_{it}$ increases, which is because this results in a reduced variance in the data on the risk scale (e.g. $\var{\mbox{SIR}_{it}=y_{it}/e_{it}}=\theta_{it}/e_{it}$), making cluster detection less demanding. \textbf{Model 1} with no smoothing component almost uniformly outperforms the remaining clustering models in terms of Rand index, with values that are higher or as high for nearly all scenarios.  This is because \textbf{Model-1} is not trying to partition the spatial variation in risk between competing model components $(\lambda_{t,Z_{it}}, \phi_{it})$, so that true cluster structure is not wrongly captured by the smoothing component. Finally, the clustering models are all conservative in terms of cluster identification, as when disease prevalence is low and the elevated risks are not that high (scenarios 2 and 4) no clusters are identified. Therefore, overall \textbf{Model-1} is the most consistent model for cluster detection, while \textbf{Model-2} is the most consistent for risk estimation.

\section{Results from the case studies}

\subsection{Modelling}
We apply four models to the Georgia and England case studies, namely the existing global smoothing models \textbf{Model-KH} and \textbf{Model-RLM}, as well as \textbf{Model-1} with no smoothing component and \textbf{Model-2} with a spatio-temporal smoothing component. The latter two are included because the simulation study showed they respectively have the best clustering and model fitting performance across the range of models proposed in this paper. Inference for each model is based on 20,000 McMC samples, which were generated following a burn-in period of 20,000 samples. In common with the simulation study convergence was assessed visually using trace plots of the McMC samples for selected parameters. The cluster models were fitted with $G=5$, although a sensitivity analysis showed changing this did not affect the results. The overall fit of each model to each data set is summarised in Table \ref{table model fit}, which displays the Deviance Information Criterion (DIC, \citealp{spiegelhalter2002}) and Log Marginal Predictive Likelihood (LMPL, \citealp{congdon2005}) statistics.

\subsection{Case study 1 - Georgia colon cancer study}
Table \ref{table model fit} shows that the DIC and LMPL statistics differ in their assessment of the best fitting model, as the DIC is minimised by \textbf{Model-KH} while the LMPL is maximised by \textbf{Model-RLM} and  \textbf{Model-2}. This is because the DIC penalises models with larger effective numbers of parameters ($p.d$), and  \textbf{Model-RLM} (and \textbf{Model-2})  has over 100 more effective parameters than  \textbf{Model-KH}.  Thus the unadjusted deviance is lower for  \textbf{Model-RLM} than  \textbf{Model-KH}, 6737.0 compared with 6796.9, suggesting a better absolute fit to the data of the former. This difference in $p.d$ is due to the relative amounts of shrinkage applied to the random effects, as for \textbf{Model-KH} the variance parameters are all less than 0.007. The exception is for the spatial main effect ($\hat{\tau}^2_{\phi}=0.180$,) which has $K$ random effects, while in \textbf{Model-RLM} $\tau^2=0.0385$ for $KN$ random effects. However, substantial smoothing of the random effects has taken place in both models, as the $p.d$ values are much smaller than the numbers of random effects in each model.\\

\begin{figure}
\centering\caption{The maps in the top row display the estimated risks for 2001 from \textbf{Model-KH}  (left panel (a)) and \textbf{Model-2} (right panel (b)), while the bottom panel (panel (c)) displays boxplots of estimated risks from \textbf{Model-2} for all years by PHD.}\label{figure georgia risk}
\begin{picture}(10,19)
\put(-3.9,1){\scalebox{0.53}{\includegraphics{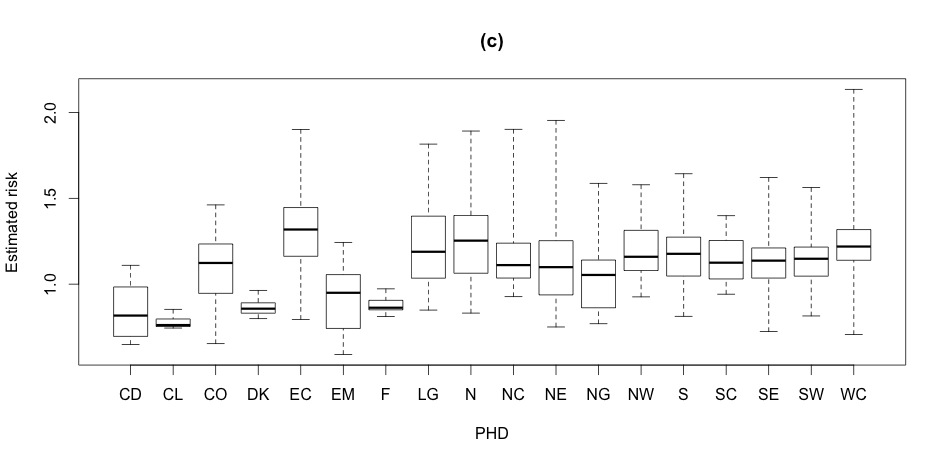}}}
\put(-3.9,10){\scalebox{0.53}{\includegraphics{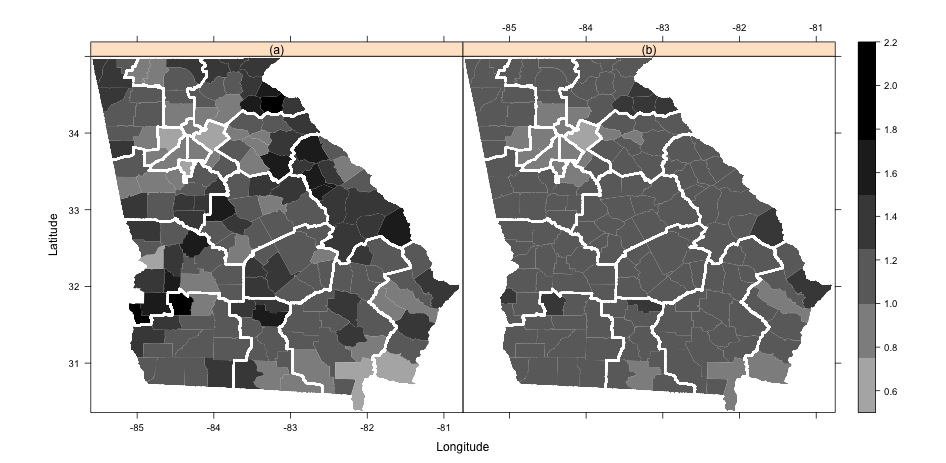}}}
\end{picture}
\end{figure}

This smoothing is evident in the top panel of Figure  \ref{figure georgia risk}, which displays the estimated risk surface $\{\theta_{it}\}$ for the first year of the study 2001 from \textbf{Model-KH} (left) and \textbf{Model-2} (right). The two models show broadly similar spatial patterns, with the main difference being that the estimates from \textbf{Model-KH} are more extreme and less smoothed than those from \textbf{Model-2}, which is due to the larger estimated spatial variance as discussed above. \textbf{Model-2} did not identify any clusters of areas at high risk for these data, as all counties were classified as being in the baseline risk class for all years. This null clustering result was also obtained when fitting \textbf{Model-1}, and suggests that the extreme SIR values are likely to be caused by a combination of small expected numbers $e_{it}$ and Poisson sampling variation. However, the figure does suggest that there are counties exhibiting slightly elevated or slightly reduced risks compared with the rest of the state, but that these are not related to PHD. An example is the group of counties in the north of the state that exhibit slightly lower risks, which cover parts of many PHD and do not respect their borders. This lack of PHD effect is confirmed by the bottom panel of Figure \ref{figure georgia risk}, which displays boxplots of estimated risks from \textbf{Model-2} by PHD for all years. The plot shows that with the exception of three PHDs that contain just one county, the variation in risks within a PHD is large, and thus the distribution of risks overlaps with the distributions from most other PHDs.

\subsection{Case study 2 - England cardiovascular study}
\begin{figure}
\centering\caption{Maps showing the posterior median risk class for 2007 quarter 2 (top left) and 2011 quarter 4 (top right) from \textbf{Model-2}, where light grey areas are in the baseline risk class while the black areas are in the elevated risk class. The Easting and Northing coordinates are in kilometres. The bottom panel displays the temporal trend in the mean of these three risk classes over time (again from \textbf{Model 2}), where the implementation of smoking ban is denoted by the vertical line.}\label{figure england risk}
\begin{picture}(10,21)
\put(-2.6,1){\scalebox{0.6}{\includegraphics{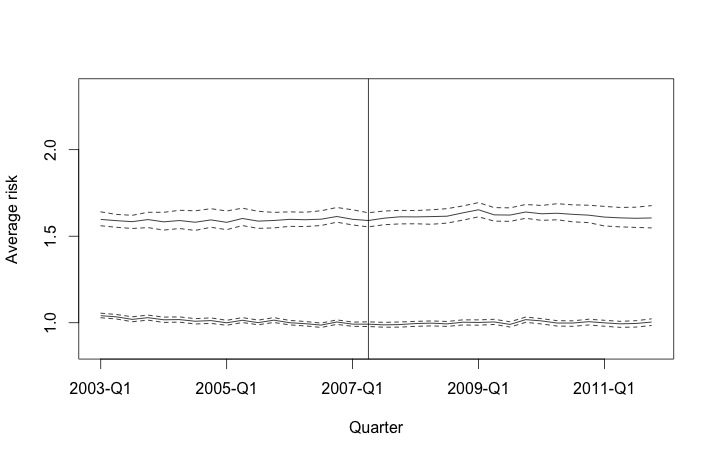}}}
\put(-4,9.5){\scalebox{0.7}{\includegraphics{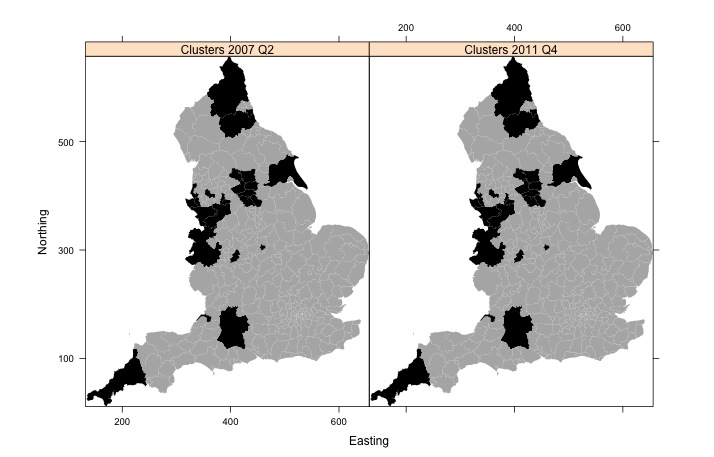}}}
\end{picture}
\end{figure}

Table \ref{table model fit} shows no clear consensus on the best fitting model, as the DIC suggests \textbf{Model-KH} followed by \textbf{Model-2}  and then \textbf{Model-RLM}, while the LMPL suggests \textbf{Model-RLM}  followed by \textbf{Model-2}  and \textbf{Model-KH}. The cluster structure estimated by  \textbf{Model-2} is largely consistent over time, with the local authorities being classified into two different risk classes for 34 out of the 36 time periods. These risk classes are a baseline class and an elevated risk class, which contain 298 and 25 areas on average over time. The allocation of local authorities to these risk classes is also largely consistent, as there is an average 99.7$\%$ agreement in class allocation between any two time periods. The risk class allocations for quarter 2 in 2007 (just before the implementation of the smoking ban) and quarter 4 in 2011 (the last time period) are displayed in the top panel of Figure \ref{figure england risk}, where areas in the baseline and elevated risk classes are shaded light grey and black respectively. The figure shows that the class structures before and after the smoking ban are identical, with areas in the elevated classes typically being in mid to northern England including the cities of Birmingham, Liverpool, Manchester and Leeds. The exceptions to this are the southern authorities of Bristol, Wiltshire and Cornwall, and this cluster structure follows the pattern of socio-economic deprivation closely.\\

The bottom panel in Figure \ref{figure england risk} displays the posterior median (solid line) and 95$\%$ credible intervals (dashed lines) for the average risk in each class (e.g. $\exp(\lambda_{t, Z_{it}})$), and the vertical line denotes the beginning of the smoking ban in July 2007. The figure shows that the baseline and elevated risk classes exhibit very little temporal trend, with average risks of 1.010 and 1.593 before the smoking ban and 0.999 and 1.619 after the ban. Computing the mean estimated risk before and after the smoking ban shows that 154 local authorities exhibit reduced risks after the ban compared to 169 that exhibit elevated risks. However, these risk differences are small, and only 6 and 4 local authorities exhibit more than a 10$\%$ decrease or increase in risk respectively.  These results tentatively suggest that the smoking ban has not yet impacted cardiovascular disease risk and clustering, although further study over a longer time period is required to make more definitive conclusions.

\section{Discussion}
This paper has proposed one of the first models for simultaneously estimating disease risk and identifying high-risk clusters, and is accompanied by freely available software, the R package \emph{CARBayesST}. Existing approaches either estimate a spatio-temporal smoothed risk surface without cluster detection, or utilise testing based paradigms to identify clusters without estimating the risk surface for the entire study region. Some work has combined these facets into a unified model in a purely spatial context, but the extension to a spatio-temporal setting throws up a number of additional modelling challenges that this paper is the first to address.\\

A number of general themes emerged from our simulation study. First, risk estimation and cluster detection can be undertaken successfully in a single model, as \textbf{Model-2}, which combines clustering and smoothing components, showed good overall performance. It outperformed two commonly used spatio-temporal smoothing models in terms of risk estimation as quantified by RMSE, and led to good overall cluster recovery. The exception to this is when the disease is rare ($e_{it}$ is small), which is because of the increased sampling variation around the raw risk estimates $\hat{\theta}_{it}=y_{it}/e_{it}$. A simplified model with $\phi_{it}=0$ performs better in terms of clustering because the two components $(\lambda_{t,Z_{it}}, \phi_{it})$ are not competing with each other to represent the same spatio-temporal variation, but this is at the expense of accuracy in terms of risk estimation. Applying a post-hoc clustering algorithm to smoothed risk surfaces performs uniformly poorly throughout, because one is trying to identify differences in risk between neighbouring areal units from a smoothed surface. Finally, setting the maximum number of risk classes $G$ to be overly large and penalising $\{ Z_{it}\}$ towards the middle class works well, as long as the penalty is not undermined by additional smoothing constraints, such as the temporal smoothing constraints considered here.\\

In common with the vast majority of the disease mapping literature our clustering models do not include covariates, so that the risk classes relate to the risk in an area and not the residual or unexplained component of risk after adjusting for the covariates. This allows us to classify groups of areas at elevated risk, so public health interventions can be appropriately targeted. However, classifying the residual risk surface after adjusting for covariates would aid in the identification of unknown etiologic covariates, as the residual class structure could be examined to look for similarity with spatially varying covariates. Covariate information could also be used as predictors for the cluster allocation model ($Z_{it}$), and both these areas will be investigated in future work.\\

Another avenue for future work is illustrated by the simulation and Georgia case studies, which show the difficulties in cluster detection and risk estimation for rare diseases. All models smoothed the extreme raw SIRs towards the null risk of one, as the spatial smoothing priors dominated the count data. Thus further work is required to develop spatio-temporal disease mapping models specifically designed for low count data $y_{it}$. Another future direction will extend the cluster models to consider multiple diseases simultaneously, which throws up additional modelling challenges. One of these key challenges will be whether and how to extend the penalty prior for $\{Z_{it}\}$ to account for between disease correlations.

\section*{Acknowledgements}
Hospital admission records from the Health and Social Care Information Centre (\emph{www.hscic.gov.uk})
were analysed at the UK Met Office by Dr Christophe Sarran to provide counts of hospital admissions for circulatory disease by local authority.

\section*{Supplementary material}
Supplementary material accompanying this paper is available, and includes exploratory analyses for the two cases studies as well as full simulation results for the second simulation study.

\bibliographystyle{chicago}
\bibliography{LeeLawson}

\begin{table}
\caption{\label{table simulation1} Table showing the root mean square error (RMSE, top panel) for the risk surface estimated by each of the six models, and the Rand index (bottom panel) which measures the clustering ability of each model.}
\centering\begin{tabular}{llrrrrrr}
\hline
\raisebox{-1.5ex}[0pt]{\textbf{Scenario}}&\raisebox{-1.5ex}[0pt]{$\mathbf{e}_{it}$}&\multicolumn{6}{c}{\textbf{Model}}\\
&&\textbf{Model-KH}&\textbf{Model-RLM}&\textbf{Model-1}&\textbf{Model-2}&\textbf{Model-3}&\textbf{Model-4}\\\hline
\textbf{RMSE}&&&&&&&\\
&[10, 30]&0.037  &0.035  &0.041  &0.036  &0.041&0.041\\
1&[90, 110]&0.030  &0.025  &0.039  &0.025  &0.031&0.032\\
&[190, 210]&0.028  &0.022  &0.040  &0.022  &0.026&0.029\\\hline

&[10, 30]&0.084  &0.123  &0.355  &0.123  &0.192&0.211\\
2&[90, 110]&0.045  &0.072  &0.060  &0.072  &0.087&0.058\\
&[190, 210]&0.037  &0.056  &0.048  &0.038   &0.037&0.034 \\\hline

&[10, 30]& 0.094&  0.167&  0.247&  0.166&  0.240&0.248\\
3&[90, 110]&0.051&  0.090&  0.058&  0.047&  0.049&0.045\\
&[190, 210]&0.042&  0.068&  0.058&  0.030&  0.036&0.040\\\hline

&[10, 30]& 0.149&  0.125&  0.240&  0.125&  0.151&0.170\\
4&[90, 110]& 0.088&  0.075&  0.046&  0.065&  0.069&0.057\\
&[190, 210]&0.067&  0.059&  0.043&  0.034&  0.035&0.032\\\hline

&[10, 30]& 0.194&  0.164&  0.169&  0.163&  0.193&0.189 \\
5&[90, 110]&0.099&  0.090&  0.049&  0.047&  0.047&0.038\\
&[190, 210]&0.072&  0.067&  0.050&  0.027&  0.035&0.034\\\hline

\textbf{Rand}&&&&&&\\
&[10, 30]&0.683&  0.704&  1.000&  1.000&  1.000&1.000\\
1&[90, 110]&0.669&  0.652&  1.000&  1.000&  1.000&1.000\\
&[190, 210]&0.707&  0.673&  1.000&  1.000&  1.000&1.000\\\hline

&[10, 30]&0.890&  0.979&  0.749&  0.726&  0.726&0.726\\
2&[90, 110]&0.925&  0.958&  0.996&  0.731&  0.851&0.981\\
&[190, 210]&0.885&  0.948&  1.000&  0.931&  0.984&1.000\\\hline

&[10, 30]&0.905&  0.945&  0.969&  0.726&  0.741&0.912\\
3&[90, 110]&0.929&  0.958&  1.000&  0.977&  0.998&1.000\\
&[190, 210]&0.899&  0.966&  0.998&  0.997&  0.998&0.998\\\hline

&[10, 30]&0.766&  0.770&  0.878&  0.878&  0.878&0.878\\
4&[90, 110]&0.932&  0.798&  0.999&  0.905&  0.929&0.958\\
&[190, 210]&0.965&  0.901&  0.998&  0.970&  0.984&0.998\\\hline

&[10, 30]&0.642&  0.712&  0982&  0.878&  0.887&0.920\\
5&[90, 110]&0.977&  0.843&  0.993&  0.970&  0.983&0.991\\
&[190, 210]&0.982&  0.929&  0.952&  0.974&  0.962&0.965\\\hline
\end{tabular}
\normalsize
\end{table}

\begin{table}
\caption{Summary of the overall fit to the data of each model, as measured by the DIC (and effective number of parameters $p.d$) and LMPL. The top panel of the table relates to the Georgia colon cancer case study, while the bottom panel relates to the England circulatory data.} \label{table model fit}
\centering\begin{tabular}{lrrrr}
\hline
&\multicolumn{4}{c}{\textbf{Model}}\\
\textbf{Metric}&\textbf{Model-KH}&\textbf{Model-RLM}&\textbf{Model-1}&\textbf{Model-2}\\\hline

\textbf{Georgia data}&&&&\\
DIC ($p.d$)&6917.9 (121.0)&6964.5 (227.5)&7702.4 (2.5)&6968.5 (227.9)\\
LMPL&-3340.2&-3271.8&-3848.5&-3272.6\\\hline

\textbf{England data}&&&&\\
DIC ($p.d$)&100,476.4 (1842.7) & 101,551.8 (4242.9)&161,004.7 (123.0)&100,817.5 (3560.5)\\
LMPL&-48,528.7&-47,146.6&-80,334.7&-47,261.2\\\hline
\end{tabular}
\end{table}

\end{document}